\documentclass[aps,pra,twocolumn,10pt]{revtex4-2}
\usepackage{amsfonts,amssymb,amsbsy,amsmath,amsthm,enumerate,verbatim}
\usepackage{color}
\usepackage{bm}
\usepackage{mathbbol}
\usepackage{braket}
\usepackage{hyperref}

\usepackage{graphicx}

\usepackage{ulem}
 
\newcommand{\rmi}{{\mathrm i}}
\newcommand{\xpo}{x_{\rm KH}}
\newcommand{\ppo}{p_{\rm KH}}

\def\beq{\begin{equation}}
\def\eeq{\end{equation}}
\def\bi{\begin{itemize}}
\def\ei{\end{itemize}}
\newcommand{\bea}{\begin{eqnarray}}
\newcommand{\eea}{\end{eqnarray}}
\newcommand{\beas}{\begin{eqnarray*}}
\newcommand{\eeas}{\end{eqnarray*}}

\begin{document}

\title{Scars of Kramers-Henneberger atoms}

\author{E. Floriani}
\email{elena.floriani@univ-amu.fr}
\affiliation{Aix Marseille Univ, Université de Toulon, CNRS, CPT, Marseille, France}

\author{J. Dubois}
\email{jonathan.dubois@sorbonne-universite.fr}
\affiliation{Sorbonne Université, CNRS, Laboratoire de Chimie Physique – Matière et Rayonnement, LCPMR, 75005 Paris, France}

\author{C. Chandre}
\email{cristel.chandre@cnrs.fr}
\affiliation{CNRS, Aix Marseille Univ, I2M, 13009 Marseille, France}

\begin{abstract}   
Electron motion in an atom driven by an intense linearly polarized laser field can exhibit a laser-dressed stable state, referred to as the Kramers-Henneberger (KH) state or KH atom. Up to now, the existence conditions of this state rely on the presence of a double well in the KH potential, obtained by averaging the motion over one period of the laser. However, the approximation involved in the averaging is largely invalid in the region of the double well structure; therefore this raises the question of its relevance for identifying signatures of these exotic states. 
Here we present a method to establish conditions for the existence of the KH atom based on a nonperturbative approach. We show that the KH atom is structured by an asymmetric periodic orbit with the same period as the laser field in a wide range of laser parameters. Its imprint is clearly visible on the wavefunction in quantum simulations. We identify the range of parameters for which this KH state is effective, corresponding to an elliptic periodic orbit. 
\end{abstract}

\maketitle

\section*{Introduction}

Subjecting neutral atoms or molecules to super-intense laser pulses revealed some counter-intuitive, intricate and fascinating phenomena, such as ionization stabilization~\cite{Geltman1974,Gersten1976,Gavrila1984,Su1990,Pont1990,Grobe1991,Kulander1991,Grochmalicki1991,Gajda1992,Eberly1993,Gavrila2002,Norman2015} and the ponderomotive acceleration of neutral atoms~\cite{Eichmann2009}, in intensity regimes where full ionization was thought to be unavoidable. A possible explanation behind these unexpected phenomena was formulated by moving to a suitable frame --using the Kramers-Henneberger (KH) coordinates~\cite{Kramers1956,Henneberger1968,Breuer1992}-- where a fast oscillation can be averaged out to derive an effective potential --called KH potential-- revealing a potential well with local minima at about one quiver radius. This potential well was hypothesized as the bedrock for the existence of some exotic laser-dressed states of the neutral atom. A KH state, also referred to as KH atom, is defined as a bound state localized in the vicinity of the local minima of the KH potential. Exploiting the physical properties of these states holds many promises as Rydberg-like atoms piloted by the laser field. However, up to now, the existence of KH atoms remains largely elusive.   

A large amount of works was produced in the early 90s, evidencing a strong enthusiasm for the perspective of manipulating these exotic atomic states. The lack of direct experimental evidence, coupled with the unrealistic values of the laser parameters (too high intensity and too small wavelength) for ionization stabilization has hindered the exploitation of the full potential of the KH atom. 

For laser parameters conventionally explored in strong field atomic physics experiments, typically in the infrared regime for intensities in the range from $10^{13}$ to $10^{16}$ W cm$^{-2}$, the approximations involved in the derivation of the KH potential are largely invalid, given that the harmonics of the Hamiltonian cannot be neglected~\cite{Morales2011}, questioning the validity of the KH atom. Some indirect signatures have been attributed to the KH atom by using photoelectron spectroscopy~\cite{Morales2011,Bray2020,Ivanov2022} for the potassium atom, and by looking at the Kerr effect response~\cite{Richter2013}. These encouraging signs raise the question of when and how KH atoms can be observed, even in a range of parameters where the KH approximation fails.

Here we address this question by looking at the periodic orbits in the corresponding classical system. We show that the KH atom corresponds (up to a symmetry) to a single elliptic or weakly hyperbolic periodic orbit mostly localized around one quiver radius. We analyze its properties as a function of the intensity of the laser and the atom under consideration and illustrate its relevance as a scar in the quantum wavefunction. 

In Sec.~\ref{sec:KHmodel}, we recall the Hamiltonian in the KH coordinates, the KH approximation and the KH potential. In Sec.~\ref{sec:classical}, we investigate the nonlinear dynamics of the KH Hamiltonian and identify the relevant periodic orbit as the classical KH state. In Sec.~\ref{sec:quantum}, we analyze the influence of the KH periodic orbit on the corresponding quantum wavefunction. 

\section{KH model, KH approximation and KH atom}
\label{sec:KHmodel}

\begin{figure}
    \centering
    \includegraphics[width=0.5\textwidth]{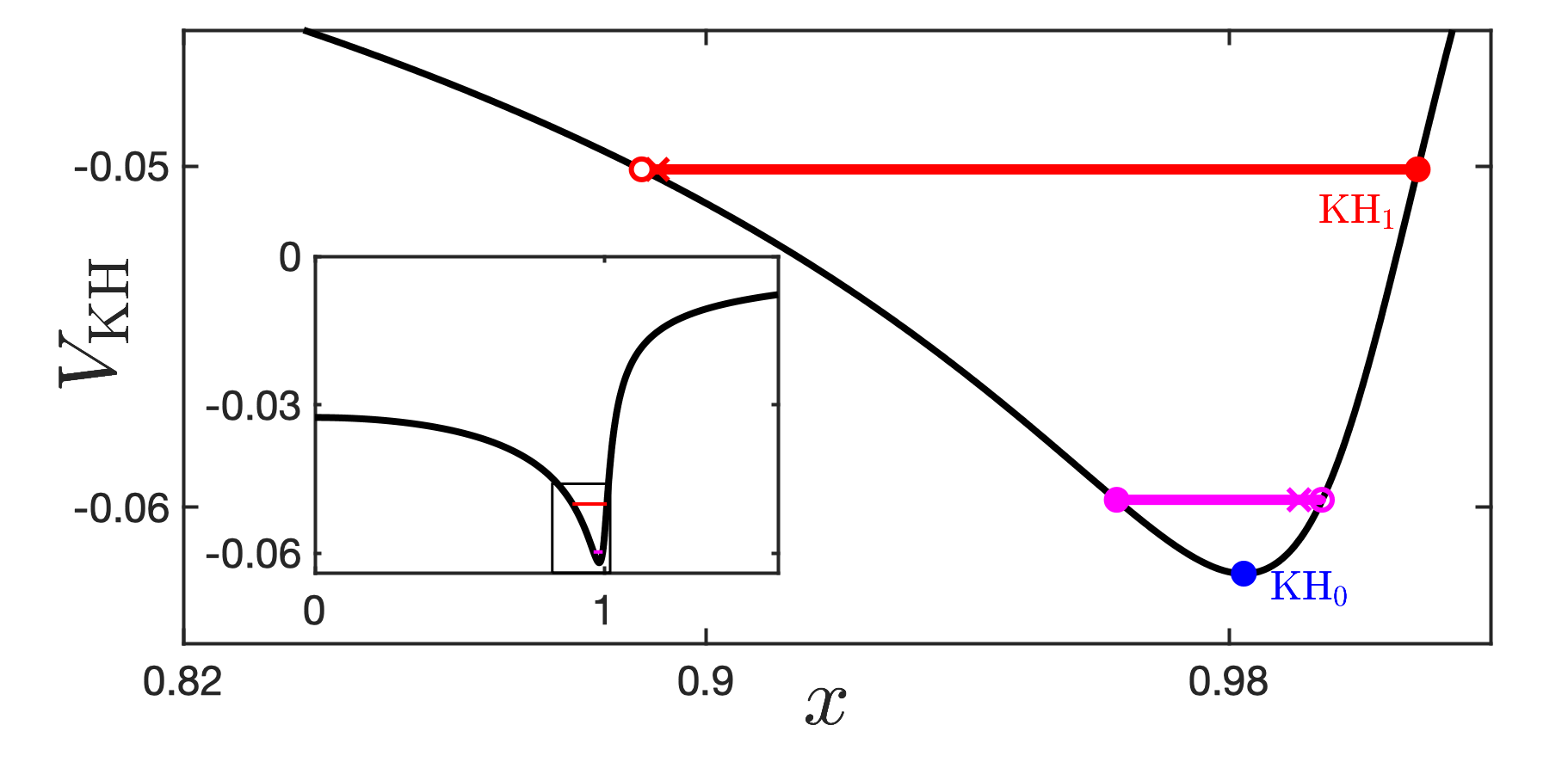}
    \caption{Effective potential $V_{\rm KH}$ given by Eq.~\eqref{eq:VKH_1D} for $I=10^{15}$ W cm$^{-2}$, $\lambda = 780$~nm and $a=1$ a.u.. The position $x$ is in units of quiver radius, the potential is in units of $U_p=E_0^2/(4\omega^2)$. Blue dot: position of the fixed point KH$_0$. Magenta line: $n=2$ resonance; magenta dots: positions of the corresponding elliptic orbits; magenta cross: position of the two symmetric hyperbolic orbits. Red line: $n=1$ resonance; red dots: positions of the corresponding elliptic orbits; red cross: position of the two symmetric hyperbolic orbits. The full red circle corresponds to the periodic orbit KH$_1$.}
    \label{fig:plotpot}
\end{figure}

In order to showcase our results, we consider the simplest argument with which the KH atom has been advocated in the literature for decades. It starts with a one-dimensional Hamiltonian model comprising a ionic potential $V$ and a linearly polarized laser field, with $E_0$ the amplitude of the electric field and $\omega$ its frequency (defined from its wavelength $\lambda$). In the dipole approximation, the Hamiltonian for the dynamics of the electron is given by
\beq
    \label{eqn:H1_1D}
    H(x_{\rm e},p_{\rm e},t) = \frac{p_{\rm e}^2}{2} +V(x_{\rm e}) + x_{\rm e} E_0 \cos \omega t  \,,
\eeq
where $V$ is a soft-Coulomb potential $V(x_{\rm e}) = -(x_{\rm e}^2+a^2)^{-1/2}$ with a softening parameter $a$. Here $x_{\rm e}$ and $p_{\rm e}$ are the classical position of the electron in phase space, or the quantum operators of position and momentum, depending on the framework. In the Kramers-Henneberger coordinates~\cite{Kramers1956,Henneberger1968}, 
\begin{subequations}
    \label{eqn:xxe}
    \begin{eqnarray}
       && x=x_{\rm e} - \frac{E_0}{\omega^2} \cos \omega t,\\
       && p=p_{\rm e} + \frac{E_0}{\omega} \sin \omega t,
    \end{eqnarray}
\end{subequations}
the Hamiltonian becomes~\cite{Kramers1956,Henneberger1968}
\beq 
    \label{eqn:H2_1D}
    H_{\rm KH}(x, p, t) = \frac{p^2}{2} + V \big( x + q  \cos \omega t \big) \,,
\eeq 
where $q=E_0/\omega^2$ is the quiver radius. In order to exhibit the local minimum of an effective potential, the dynamics is averaged over one period (linked with the fast oscillations of the field), so that the Hamiltonian becomes 
\beq 
    \label{E2_1D}
    \langle H_{\rm KH} \rangle(x, p) = \frac{p^2}{2}+ V_{\rm KH}(x) \,,
\eeq
where $V_{\rm KH}$ is the Kramers-Henneberger potential given by
\beq 
    \label{eq:VKH_1D}
    V_{\rm KH} (x) = \frac{1}{2\pi}\int_0^{2\pi} V \big( x + q  \cos \phi \big) \,{\rm d}\phi \,.
\eeq
Consequently, if the KH approximation is valid, meaning that the fast oscillations can be averaged out, the KH effective Hamiltonian $\langle H_{\rm KH} \rangle$ is time independent, and the classical dynamics evolves on a constant energy surface $E_{\rm KH} = \langle H_{\rm KH} \rangle(x, p)$. 
Figure~\ref{fig:plotpot} shows an example of KH effective potential~\eqref{eq:VKH_1D}. In a wide range of laser parameters $V_{\rm KH}$ exhibits two minima, located at $\pm\, x_{\rm KH}$, just about one quiver radius away from the parent ion, corresponding to two fixed points in the phase space of the KH effective Hamiltonian $\langle H_{\rm KH} \rangle$. 

The bound eigenstates of $\langle H_{\rm KH} \rangle$, denoted $\ket{ \psi_n}$, exhibit probability density maxima around $\pm x_{\rm KH}$. In principle, the wavefunction may stabilize in these potential wells as a consequence of the existence of the bound states $\ket{ \psi_n}$.
However, the elegance of the argument advocating the existence of these laser-dressed states has not been matched with some conclusive evidence. Neglecting the terms resulting from averaging out the fast oscillations is far from an obvious step, and depends significantly on the values of the parameters of the laser and the atom, and the region in phase space where this approximation is performed. 
In other words, is the existence of $\ket{\psi_n}$ a sufficient condition for stabilization?
To address this question, we introduce the Hamiltonian
\begin{equation}
    \label{eqn:He}
    H_{\varepsilon}(x,p,t) = \varepsilon H_{\rm KH}(x,p,t) + (1-\varepsilon)\langle H_{\rm KH} \rangle(x,p) \;,
\end{equation}
with $\varepsilon$ a parameter that allows to move from a time evolution governed by $\langle H_{\rm KH} \rangle$ ($\varepsilon=0$) to $H_{\rm KH}$ ($\varepsilon=1$). 
For $\varepsilon = 0$, the population of $\ket{\psi_n}$ is conserved in time. As a consequence, the probability density is localized around $\pm x_{\rm KH}$.
For $\varepsilon \neq 0$, the population of those states evolves in time. However, there is no stability condition for the population of $\ket{\psi_n}$ and no clear evidence about the role or relevance of this eigenbasis for $\varepsilon = 1$.
Moreover, for most atoms and for laser fields in the near-infrared regime and in the $10^{13}$ to $10^{16}$ W cm$^{-2}$ intensity range --nowadays routinely considered in experiments--, the KH approximation is clearly not valid~\cite{Floriani2022_PhyD} in a region around one quiver radius away from the ionic core.
Can we then still have a KH atom even if the KH approximation is not valid? In other terms, if the KH potential does not structure the dynamics, what are the objects in phase space organizing the dynamical processes? We address this question by first considering the nonlinear dynamics associated with Hamiltonian~\eqref{eqn:He}.

\section{KH atom as a periodic orbit}  
\label{sec:classical}

\subsection{Finding KH state(s) by a continuation method}

In the neighborhood of the KH fixed point (for $\varepsilon=0$), denoted KH$_0$ in what follows, there is a harmonic motion with frequency $\omega_{\rm KH}=\sqrt{V_{\rm KH}''(x_{\rm KH})}$. Still for $\varepsilon=0$, away from the fixed point (for higher effective energy $E$), the phase space is densely filled with periodic orbits of increasing periods, denoted $2\pi/\Omega(E)$, bouncing back and forth on the walls of the potential $V_{\rm KH}$, as illustrated in Fig.~\ref{fig:plotpot}. When the time dependence of the field is taken into account, the fixed point KH$_0$ of $\langle H_{\rm KH} \rangle$ is in fact a periodic orbit of period $T=2\pi/\omega$. This periodic orbit is stable and elliptic since its location corresponds to a minimum of the effective potential energy. The motion bouncing back and forth on the potential walls corresponds now to two-dimensional invariant tori with the two frequencies $(\Omega(E), \omega)$. When the two frequencies are commensurate, i.e., there exists $(n,m)\in {\mathbb N}^2$ such that $n\omega-m\Omega(E)=0$, the invariant tori are densely filled with periodic orbits (with period $2\pi m /\omega$). Most of these periodic orbits will be broken as soon as the time dependence is turned on (i.e., for $\varepsilon>0$). However, some periodic motion originating from the KH potential persists, and these will be referred to as the KH periodic orbits. The question is then to know whether some of the KH periodic orbits persist up to the full Hamiltonian $H_{\rm KH}$ by increasing gradually $\varepsilon$ away from 0 and up to 1 in Hamiltonian~\eqref{eqn:He}, i.e., when the KH approximation is not applied. The quantity $1-\varepsilon$ corresponds to the degree of consideration of the KH approximation. As $\varepsilon$ increases from 0, the approximation is gradually turned off. We follow the destiny of the KH periodic orbits for increasing $\varepsilon$~: We start with a situation where the KH periodic orbits exist for $\varepsilon=0$ (i.e., when $V_{\rm KH}$ has a local minimum), and increase $\varepsilon$. If the KH periodic orbits can be followed up to $\varepsilon=1$, this corresponds to a situation where the KH atom exists. We restrict the study to the case $m=1$, i.e., $\Omega(E)=n\omega$, corresponding to periodic orbits with the same period as the laser field. The shortest periodic orbits are likely to be the most stable ones, and hence with a long-lasting impact on the dynamics. 

In Fig.~\ref{fig:plotpot}, some surviving KH periodic orbits for $\varepsilon=0^+$ are represented by circles and crosses. The periodic orbits which turn out to be of crucial importance for the destruction of the KH$_0$ state correspond to the resonant motion with the laser field, in particular, the periodic orbits of $\langle H_{\rm KH} \rangle$ with frequency $\Omega(E)=n\omega$ where $n\in \mathbb{N}^*$, denoted KH$_n$ in what follows. For our choice of the parameters $I$, $\lambda$ and $a$, there are two such resonances, whose energy levels are represented in Fig.~\ref{fig:plotpot} as a red line (the $n=1$ resonance) and a magenta line (the $n=2$ resonance).
We have followed the positions and residues~\cite{Greene1979} of the corresponding periodic orbits for increasing $\varepsilon$, and observed that the KH$_0$ state disappears by colliding with a periodic orbit relative to the closest resonance in phase space, through what is referred to as a saddle-node bifurcation.  The closest resonance is the one with $n=\lfloor \omega_{\rm KH}/\omega \rfloor$, and $n=2$ in this particular case. 

\begin{figure*}
    \centering
    \includegraphics[width=0.8\textwidth]{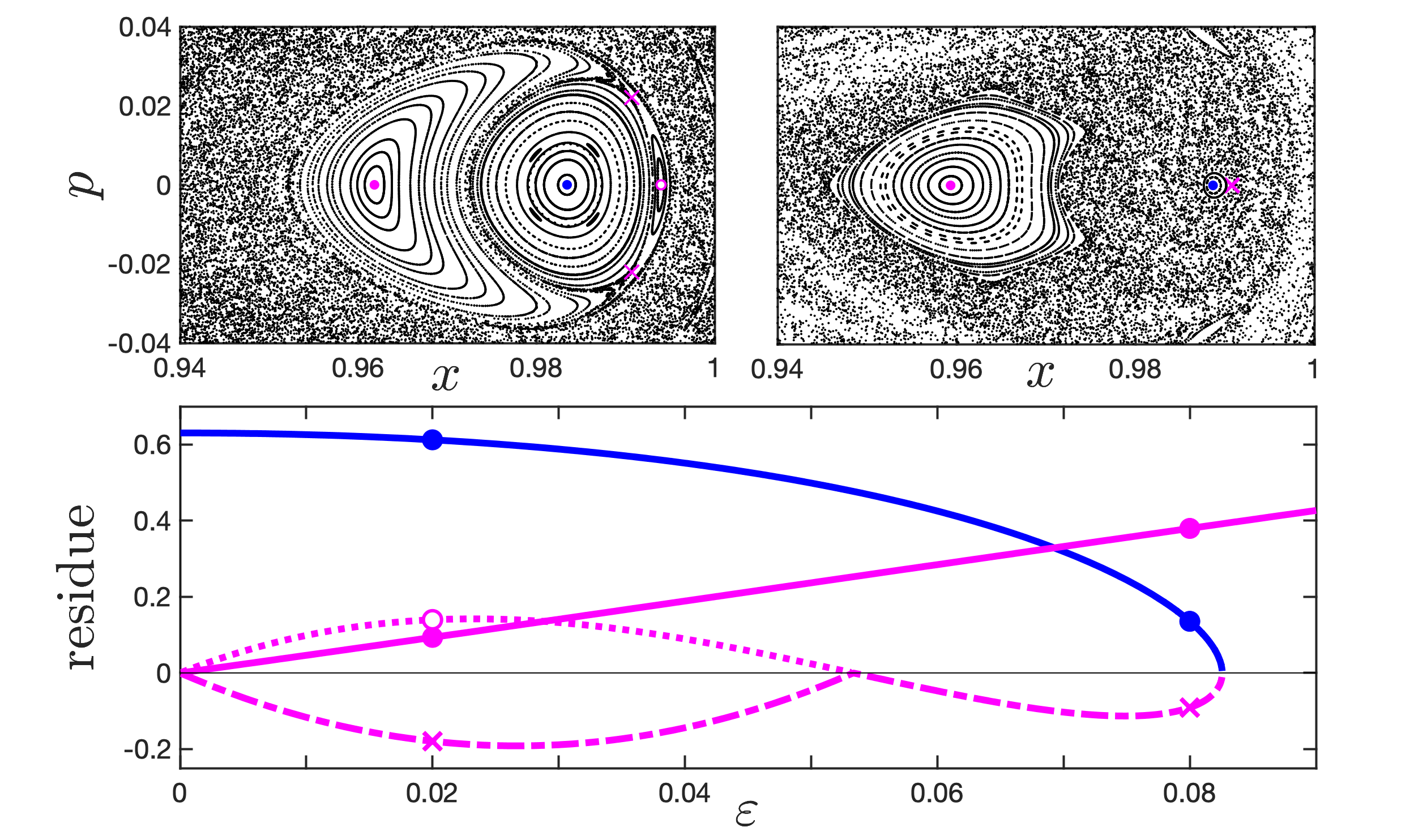}
    \caption{Upper left panel~: Poincaré section for Hamiltonian~\eqref{eqn:He} at $\varepsilon=0.02$. Upper right panel~: Poincaré section for Hamiltonian~\eqref{eqn:He} at $\varepsilon=0.08$. The positions $x$ are in units of quiver radius, the momenta $p$ are in units of $E_0/\omega$. Lower panel~: evolution of the periodic orbit residues~\cite{Greene1979} with $\varepsilon$. The symbols for the periodic orbits are the same as in Fig.~\ref{fig:plotpot}. The parameters are $I=10^{15}$ W cm$^{-2}$, $\lambda = 780$~nm and $a=1$ a.u..}
    \label{fig:poinc_res}
\end{figure*}

Figure~\ref{fig:poinc_res} illustrates the typical mechanism through which KH$_0$ disappears. The upper left panel shows a Poincaré section for Hamiltonian~\eqref{eqn:He} with $\varepsilon=0.02$, a value such that all the periodic orbits born out of the KH fixed point and the $n=2$ resonance still exist (i.e., still in the regime where the KH approximation largely applies). Their residues~\cite{Greene1979}, measuring their stability, appear in the lower panel. The blue dot labels the elliptic orbit representing the KH$_0$ state; the two magenta dots label the two elliptic orbits born out of the $n=2$ resonance; the magenta crosses label the two hyperbolic orbits born out of the resonance, which have the same $x$ position and momenta $p$ of opposite signs.

The residue curves show that at $\varepsilon\simeq 0.053$ the elliptic orbit labelled by the full magenta circle merges with the two hyperbolic orbits, to give a single hyperbolic orbit which is now symmetric around $p=0$. As $\varepsilon$ increases the surviving hyperbolic orbit and the KH$_0$ state get closer, until they collide at $\varepsilon\simeq 0.082$, so well before the KH approximation is released. The upper right panel of Fig.~\ref{fig:poinc_res} illustrates this, showing a Poincaré section for Hamiltonian~\eqref{eqn:He} with $\varepsilon=0.08$, just before the KH$_0$ state and the hyperbolic orbit disappear. Therefore, the mechanism through which the KH$_0$ orbit disappears is a collision with a periodic orbit associated with the KH potential resonant with the laser field, as a saddle-node bifurcation of periodic orbits. 
\begin{figure}
    \centering
    \includegraphics[width=0.5\textwidth]{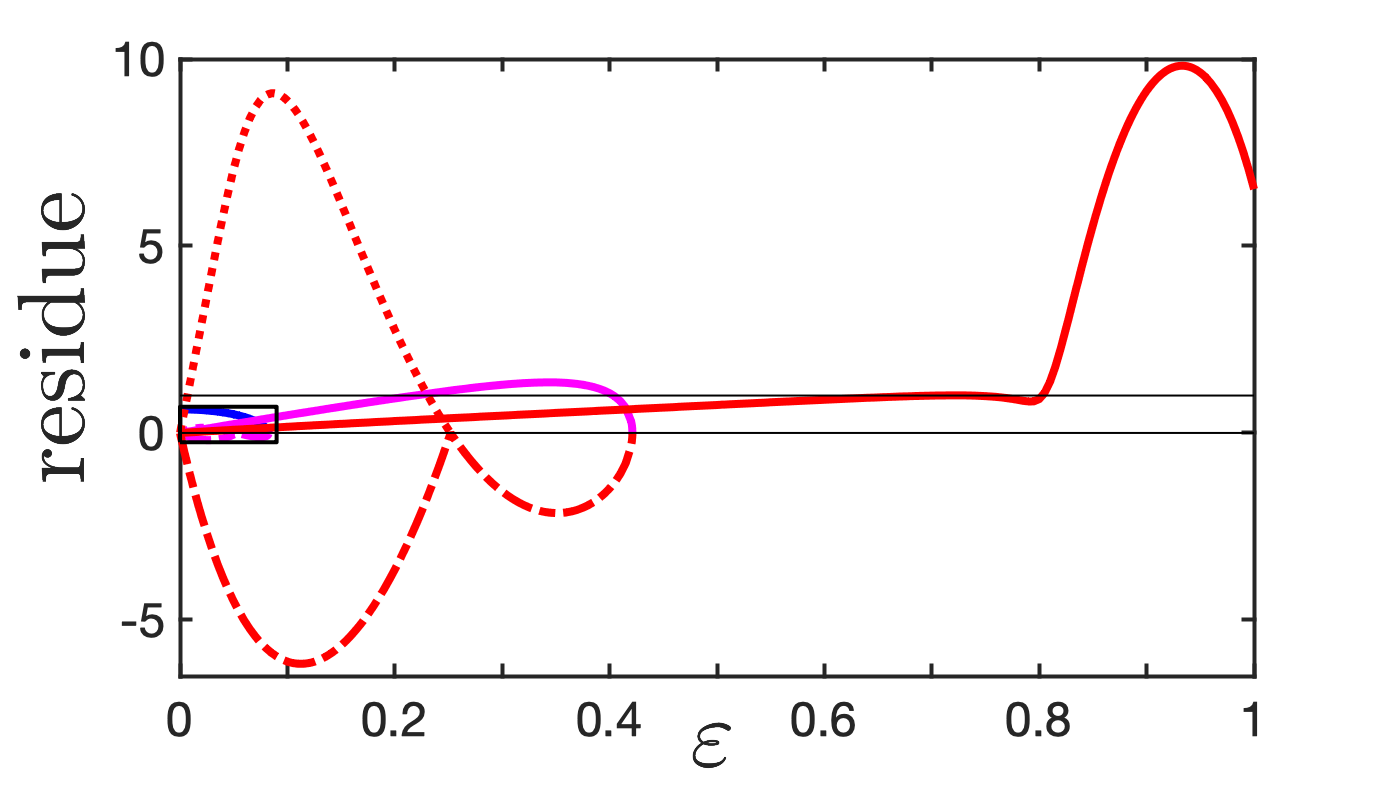}
    \caption{Evolution of the residues~\cite{Greene1979} of all the periodic orbits (KH$_0$ and $n=2$ and $n=1$ resonances) for Hamiltonian~\eqref{eqn:He}. The symbols for the periodic orbits are the same as in Fig.~\ref{fig:plotpot}. The parameters are $I=10^{15}$ W cm$^{-2}$, $\lambda = 780$~nm and $a=1$ a.u.. The black frame on the left-hand side corresponds to the limits of Fig.~\ref{fig:poinc_res}.}   \label{fig:all_res}
\end{figure}
We have carried out a similar analysis of phase space structures and confirm the generality of the scenario, a saddle-node bifurcation between the KH$_0$ state and a resonant state with $n=\lfloor \omega_{\rm KH}/\omega \rfloor$. For all the cases where this resonance exists, the only periodic orbit surviving in the dynamics of Hamiltonian~\eqref{eqn:H2_1D} is KH$_1$, issued from the $n=1$ resonance. The reason is that each resonance generates an even number of periodic orbits. After the saddle-node bifurcation that eliminates KH$_0$, one orbit issued from the $n=\lfloor \omega_{\rm KH}/\omega \rfloor$ resonance is left. This orbit is in turn destroyed by a resonant state of order $n-1$, while one state of order $n-1$ survives, and will be destroyed by a resonant state of order $n-2$. So, it is one orbit of the lowest order $n=1$, i.e., KH$_1$, which survives this process, as it is illustrated in Fig.~\ref{fig:all_res} for the case under study.

For values of the parameters such that there are no resonances with the laser field, it is the KH$_0$ state who survives in the dynamics of Hamiltonian~\eqref{eqn:H2_1D}, since there are no resonant orbits that can destroy it through the saddle-node bifurcation depicted in Fig.~\ref{fig:poinc_res}. In summary, only KH$_0$ or KH$_1$ persist up to $\varepsilon=1$. 
\begin{figure}
    \centering
    \includegraphics[width=0.5\textwidth]{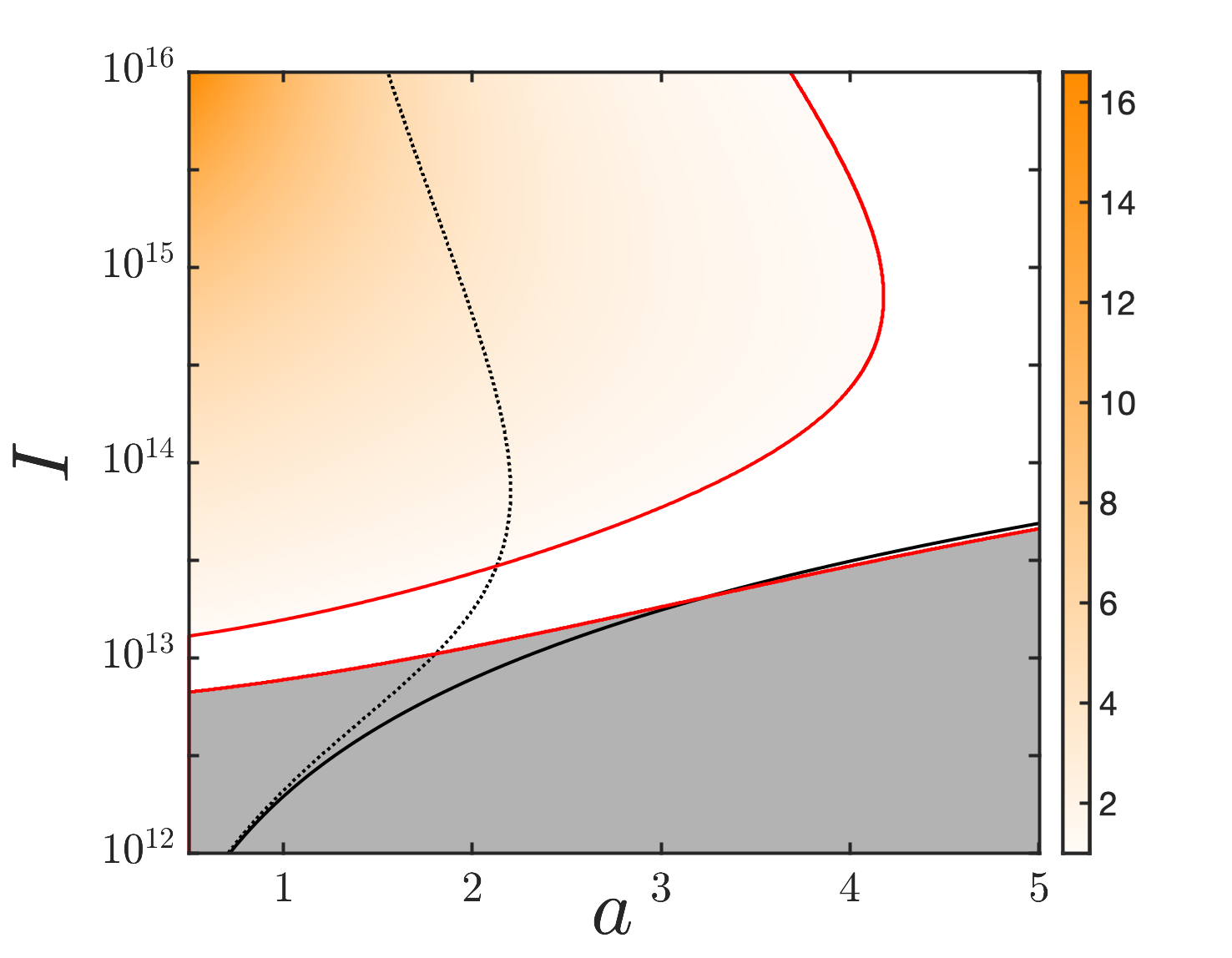}
    \caption{Domain of existence of the KH periodic orbit for Hamiltonian~\eqref{eqn:H2_1D} in parameter space $(a,I)$ for $\lambda = 780$~nm. The orange scale corresponds to the value of the residue~\cite{Greene1979} of the periodic orbit when hyperbolic. In the white region (delimited by the two red lines), the KH atom is elliptic. The region on the right-hand side of the dotted black line is such that $\omega_{\rm KH}<\omega$, and corresponds to the KH$_0$ periodic orbit; the region on the left-hand side of the dotted black line corresponds to the KH$_1$ periodic orbit ($\omega_{\rm KH}>\omega$). The gray region is the one where the KH state does not exist. Below the black solid line, the potential has no local minima at nonzero values of $x$. The values of $I$ are in W cm$^{-2}$ and the values of $a$ are in atomic units. 
    }
    \label{fig:KHstate}
\end{figure}

\subsection{For which intensities are there KH atoms?} 

We scan the parameters $I$ and $a$ and we fix the laser wavelength at $\lambda=780$~nm. The question we address is for which values of $(a,I)$ the dynamics leads to a KH atom as a continuation of the KH periodic orbit born out of the local minimum of the KH potential. 

From Fig.~\ref{fig:KHstate}, we conclude that there is a single KH periodic orbit, whether it is linked to KH$_0$ or KH$_1$ in the KH approximation (respectively, regions at the right or left of the dotted black line in Fig.~\ref{fig:KHstate}, which corresponds to $\omega_{\rm KH}=\omega$).
Clearly, both states lead to the same periodic orbit. 

For small intensities or small ionization potential (or equivalently large $a$, e.g., $a\gtrsim 4.2$ for $\lambda=780$ nm), this KH periodic orbit is elliptic (the white region in Fig.~\ref{fig:KHstate}, delimited by red lines), otherwise, it is weakly hyperbolic (the orange region in Fig.~\ref{fig:KHstate} represents the values of the residue~\cite{Greene1979} associated with the KH periodic orbit). 

For any $a$, there is a value of laser intensity below which the KH state does not exist anymore (the gray region in Fig.~\ref{fig:KHstate}). The KH state disappears due to a bifurcation through which the two asymmetric KH orbits born from the right and left wells of the KH potential coalesce to give a single symmetric orbit.

The location $x_{\rm KH}$ of the local minimum of the KH potential depends only on the ratio of the softening parameter to the quiver radius $a/q \propto a/\sqrt{I}$. The critical value for the existence of $x_{\rm KH} \neq 0$ is $a/q\simeq 0.46$ (the black solid line in Fig.~\ref{fig:KHstate}). Figure~\ref{fig:KHstate} shows that the presence of local minima of the potential $V_{\rm KH}$ (for laser intensities $I> a^2 \omega^4 \times 6.03\cdot 10^{18}$W~cm$^{-2}$) is neither a necessary nor a sufficient condition for the existence of the KH state.

\begin{figure*}
    \includegraphics[width=\textwidth]{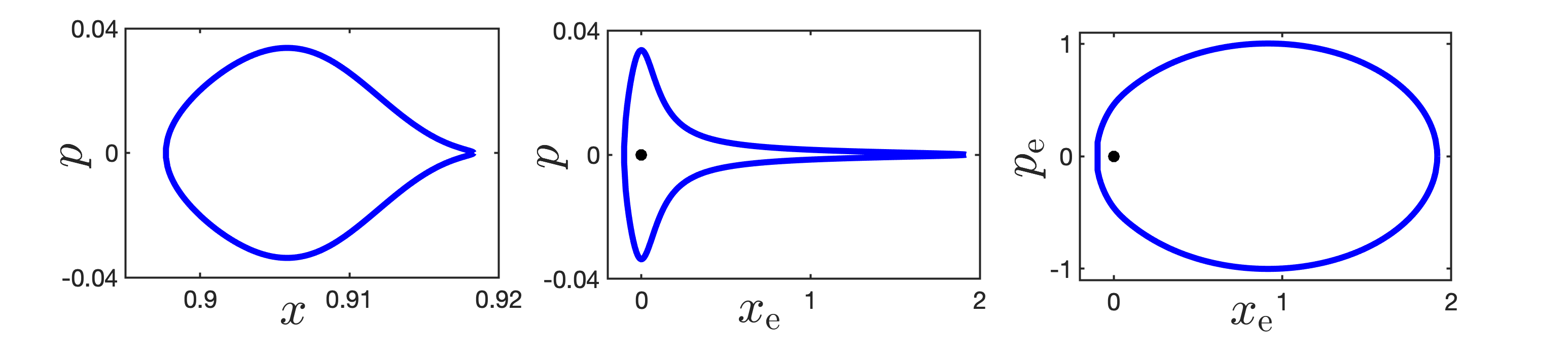}
    \centering
    \caption{The KH periodic orbit for $I=10^{15}$~W~cm$^{-2}$, $\lambda = 780$~nm, $a=5$ a.u.\ in the KH frame (left panel), semi-KH frame (middle panel) and electron coordinates (right panel). The positions $x$, $x_{\rm e}$ are in units of quiver radius $q$, and the momenta $p$, $p_{\rm e}$ are in units of $E_0/\omega$. The changes of coordinates are given by Eq.~\eqref{eqn:xxe}. The black dot gives the position of the ionic core (on the middle and right panels).}
    \label{fig:KHorbit}
\end{figure*}

\begin{figure}
    \centering
    \includegraphics[width=.5\textwidth]{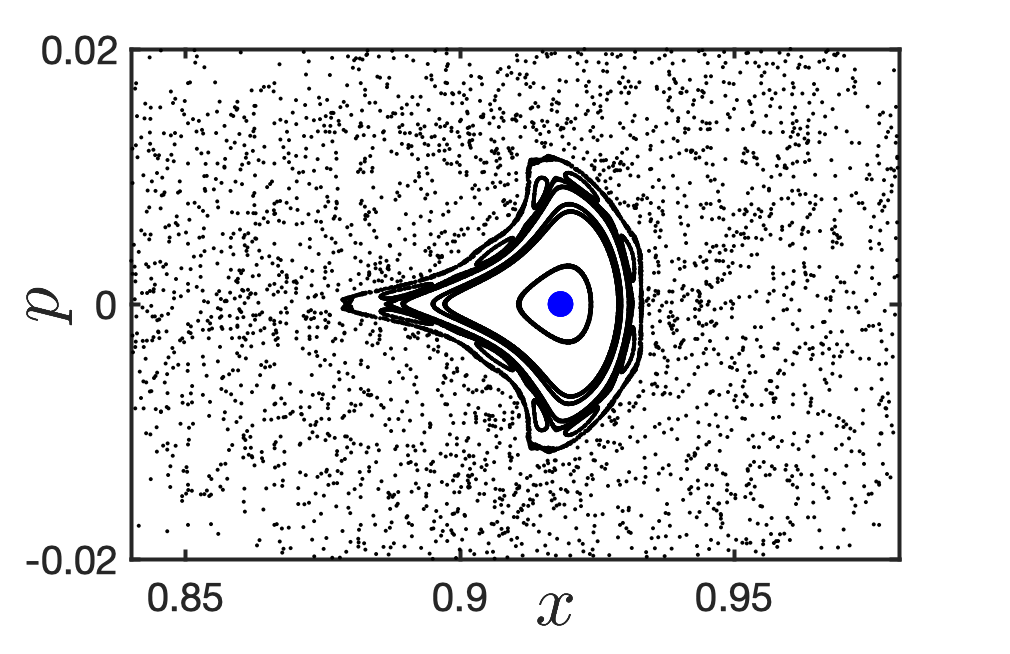}
    \caption{Poincar\'e section for Hamiltonian~\eqref{eqn:H2_1D} around the KH periodic orbit for a laser wavelength of 780~nm, $I=10^{15}$ W cm$^{-2}$ and $a=5$ a.u.. The position of the elliptic KH periodic orbit is indicated by a blue dot. The position $x$ is in units of quiver radius $q$, and the momentum $p$ is in units of $E_0/\omega$.}
    \label{fig:PS5}
\end{figure}

Figure~\ref{fig:KHorbit} represents the elliptic KH periodic orbit for $I=10^{15}$~W~cm$^{-2}$, $\lambda = 780$~nm and $a=5$ a.u.. The orbit is indeed localized around one quiver radius if expressed in the KH coordinates (left panel). Going back to the electron coordinates $(x_{\rm e}, p_{\rm e})$ of Hamiltonian \eqref{eqn:H1_1D}, the orbit is reminiscent of a Rydberg state (right panel). A Poincar\'e section of Hamiltonian~\eqref{eqn:H2_1D} close to the position of the elliptic KH periodic orbit is represented in Fig.~\ref{fig:PS5}. It shows the rather small extent (in position and momentum) of the elliptic region around this orbit.  

The KH atom exists for a wide range of experimentally accessible intensities and atoms, exemplifying its robustness. In addition, for most of the parameters, the KH atom is elliptic or very weakly hyperbolic, signifying that its effect might be clearly visible on the wavefunction, most likely for the entire duration of the laser pulse. These elements make the KH atom an ideal candidate for a laser-driven Rydberg state, provided it is explored in the relevant range of parameters. 
Here we have used the KH potential as a guide to find the relevant periodic orbit leading to the KH state. However, given that the KH approximation is not valid, the link between the KH potential and the true dynamics is tenuous. For instance, as we have discussed, there are situations where the KH potential has a local minimum but there is no KH state, and other situations where there is a KH state but no minimum of the KH potential. 

\section{Signature of the KH periodic orbit in quantum simulations} 
\label{sec:quantum}

\begin{figure*}
    \centering
    \includegraphics[width=.9\textwidth]{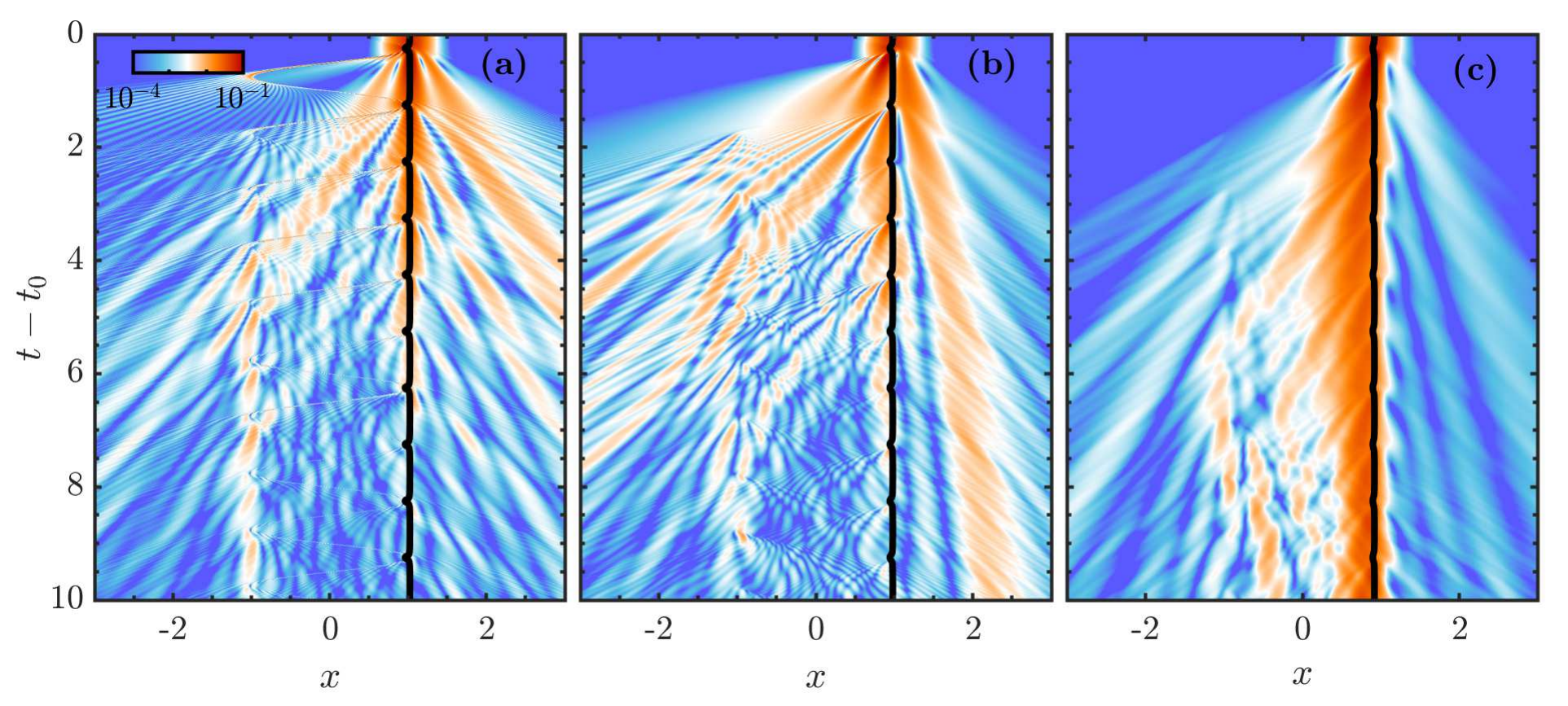}
    \caption{Probability density $\vert \psi(x,t)\vert^2$ as a function of the position $x$ and time $t$ for a KH state sculpted by the periodic orbit (black lines) for a laser wavelength of 780~nm, $t_0=1/4$, $I=10^{15}$ W cm$^{-2}$ and (a) $a=1$ a.u., (b) $a=2.5$ a.u. and (c) $a=5$ a.u.. The position $x$ is in units of quiver radius $q$, and times $t$ and $t_0$ are in units of laser cycles (l.c.).}
    \label{fig:psi}
\end{figure*}

The predictions obtained from the classical analysis are tested on the corresponding quantum system. 
Figure~\ref{fig:psi} shows the probability density as a function of the position and time obtained from the solution of the time-dependent Schr\"{o}dinger equation 
\begin{equation}
    \rmi \partial_t \psi (x,t) = \hat{H}_{\rm KH} (t) \psi (x,t) ,
\end{equation}
with $\hat{H}_{\rm KH} (t) = H_{\rm KH}(x,-\rmi \partial_x ,t)$. 
The wavefunction is initiated as a coherent state centered around the KH periodic orbit. The position and momentum of the KH periodic orbit at time $t$ are denoted $(\xpo (t),\ppo (t))$. The wavefunction is therefore initiated as 
\begin{equation}
    \psi (x,t_0) = \phi_{(\xpo (t_0),\ppo (t_0))}(x) , 
\end{equation}
where 
\begin{equation}
    \label{eq:coherent_state}
    \phi_{(x_0,p_0)}(x) = \dfrac{1}{(\pi \sigma^2)^{1/4}} \exp \left( - \dfrac{(x-x_0)^2}{2 \sigma^2} + \rmi x p_0 \right) ,
\end{equation}
represents a coherent state of mean position $x_0$ and mean momentum $p_0$.
We use a standard deviation $\sigma = 10$ a.u. corresponding to roughly 0.2 times the quiver radius for a laser intensity $I=10^{15}$ W cm$^{-2}$ and wavelength $780$ nm for all the observables we compute next.
The quantum simulations are started at time $t_0 = \pi/(2\omega)$ such that the laser electric field is zero at the beginning of the simulation. 
The wavefunction is propagated on a symmetric grid of step size 0.2 and $2^{14}$ points using a second-order split-operator method~\cite{Bandrauk1991}.
The three panels of Fig.~\ref{fig:psi} show the probability density function as a function of the position and time for different softening parameters.
In all panels, we observe that a part of the wavefunction remains localized around the KH periodic orbit for all times. 
The wavefunction is sculpted by the KH periodic orbit.
However, in panels (a) and (b), we observe that the amount of wavefunction rather quickly fades away from the (weakly) hyperbolic periodic orbits. Conversely, in panel (c), it remains highly localized around the elliptic periodic orbit.

\begin{figure}
    \centering
    \includegraphics[width=.45\textwidth]{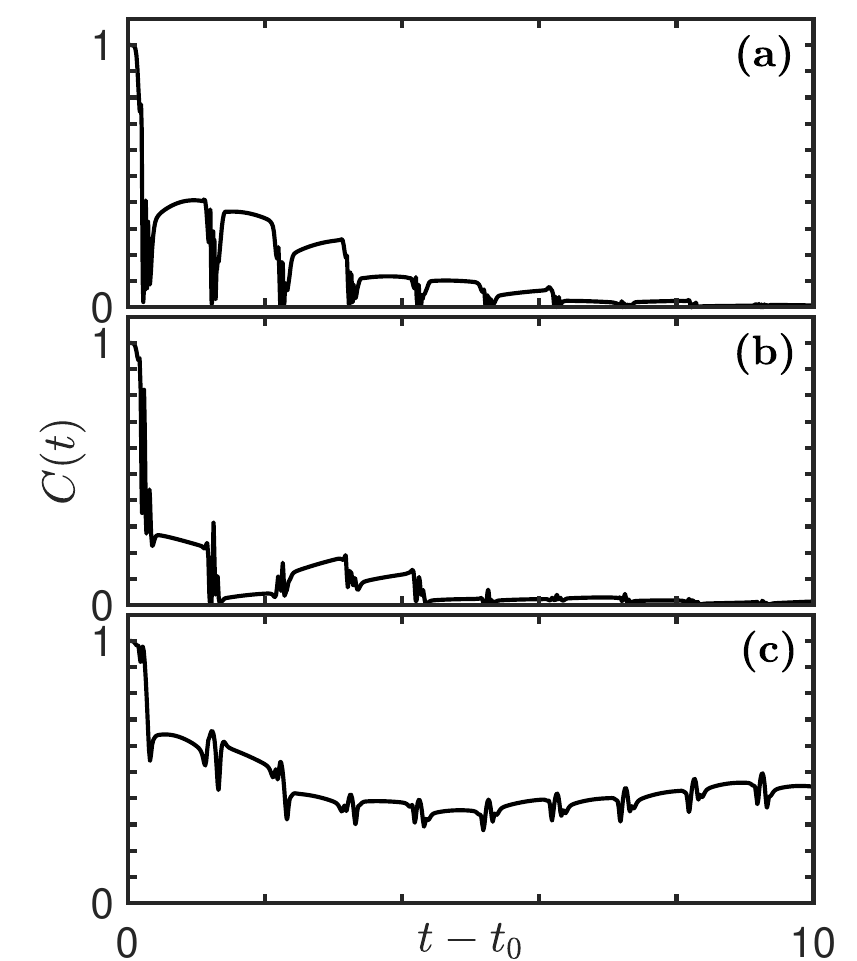}
    \caption{Correlation function $C(t)$ given by Eq.~\eqref{eqn:Ct} as a function of time $t$ for the same parameters as in Fig.~\ref{fig:psi}, i.e., for a laser wavelength of 780~nm, $I=10^{15}$ W cm$^{-2}$ and (a) $a=1$ a.u., (b) $a=2.5$ a.u.\ and (c) $a=5$ a.u.. The correlation $C(t)$ is in atomic units, and times $t$ and $t_0=1/4$ are in units of laser cycles.}
    \label{fig:Ct}
\end{figure}

\begin{figure}
    \centering
    \includegraphics[width=.45\textwidth]{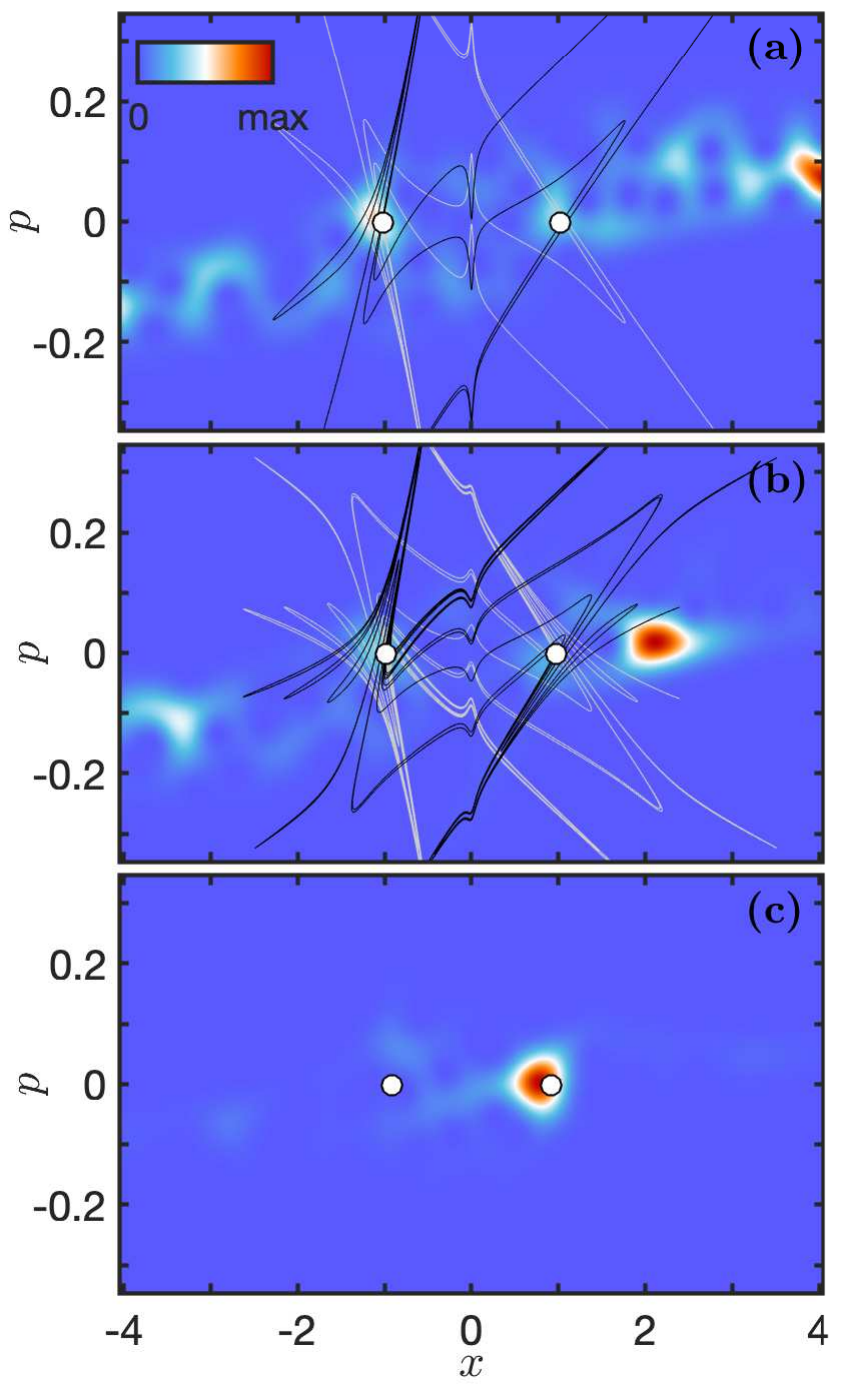}
    \caption{Husimi representation of the wavefunction in phase space, $Q(x,p,t)$ given by Eq.~\eqref{eq:Husimi}, at time $t=t_0+8$ l.c.\ for the same parameters as in Fig.~\ref{fig:psi}, i.e. for a laser wavelength of 780~nm, $t_0=1/4$ l.c., $I=10^{15}$ W cm$^{-2}$ and (a) $a=1$ a.u., (b) $a=2.5$ a.u.\ and (c) $a=5$ a.u.. The black and gray lines are the stable and unstable manifolds, respectively, of the KH periodic orbit indicated by white dots. The position $x$ is in units of quiver radius $q$, and the momentum $p$ is in units of $E_0/\omega$.}
    \label{fig:Husimi}
\end{figure}

In order to get a more quantitative insight into the localization of the wavefunction around the KH periodic orbit, we compute the correlation function between the wavefunction and a coherent state given by Eq.~\eqref{eq:coherent_state}, centered around the KH periodic orbit; the explicit expression of the correlation function is
\begin{equation}
    \label{eqn:Ct}
    C(t) = \left| \int \phi_{(\xpo (t),\ppo (t))}^{\ast} (x) \: \psi(x,t) \;  \mathrm{d} x \right|^2 .
\end{equation}
The values of $C(t)$ are shown for each case in Fig.~\ref{fig:Ct} as a function of time. In all panels, we notice that $C(t_0) = 1$.  
In panels (a) and (b), we observe that $C(t)$ quickly drops off in the first periods of the simulation, so that $C(t) < 0.05$ for $t-t_0>5$.
On the contrary, in panel (c), we observe that $C(t)$ decreases slowly with time, so that after ten laser cycles, $C(t) \approx 0.45$.
Thus, the wavefunction remains close to the classical KH periodic orbit if it is elliptic. This demonstrates that elliptic KH periodic orbits act as a backbone for the wavefunction dynamics. If the KH periodic orbit is hyperbolic, the wavefunction only remains localized around it for a few laser cycles, then spreads away from it rather quickly. 
However, in contrast to elliptic periodic orbits, hyperbolic ones have invariant manifolds that structure phase space. We study the influence of the invariant manifolds of the hyperbolic KH periodic orbit on the wavefunction by comparing the classical and quantum views of the dynamics in phase space. Quantum mechanically, we use the Husimi representation 
\begin{equation}
    \label{eq:Husimi}
    Q(x,p,t) = \left| \int \phi^{\ast}_{(x,p)} (y) \, \psi (y,t) \; {\rm d} y \right|^2 ,
\end{equation}
corresponding to the overlap between the wavefunction $\psi (x,t)$ and a coherent state given by Eq.~\eqref{eq:coherent_state} centered at $(x,p)$~\cite{Harriman1993}. The integral in Eq.~\eqref{eq:Husimi} can be written in the form of a convolution product and can therefore be efficiently computed using fast Fourier transforms.
At time $t=t_0$, $Q(x,p,t_0)$ is centered around the KH periodic orbit with positive $x$.
As time evolves, $Q(x,p,t)$ follows the periodic orbit motion and leaks slowly out of its surrounding region as indicated by the correlation functions in Fig.~\ref{fig:Ct}. 
An animated gif for the time evolution of the Husimi function is provided in the Supplemental Material of this article~\cite{SM}. During the time evolution, for all cases, we observe that the Husimi representation of the wavefunctions follows the motion of the periodic orbits with high fidelity. The wavefunction spreads throughout phase space for the hyperbolic case, and remains rather localized around the periodic orbit for the elliptic case. This is also what can be seen from Figs.~\ref{fig:psi} and~\ref{fig:Ct}.
Figure~\ref{fig:Husimi} shows $Q(x,p,t)$ at time $t-t_0=8$ laser cycles for the same parameters as in Fig.~\ref{fig:psi}.
In the three panels of Fig.~\ref{fig:Husimi}, we observe that a part of the electron remains localized in phase space around the KH periodic orbit. 
In Fig.~\ref{fig:Husimi}(c), for which the KH periodic orbit is elliptic, we observe that the electron remains in its neighborhood.
In Figs.~\ref{fig:Husimi}(a) and~\ref{fig:Husimi}(b), the stable and unstable manifolds of the KH periodic orbits are indicated by gray and black lines, respectively. In this case, classically and quantum mechanically, the electron escapes the neighborhood of the hyperbolic KH periodic orbit through its invariant manifolds. We observe that local maxima of $Q(x,p,t)$ are located around the intersections of the stable and unstable manifolds. The wavepackets are driven back and forth from the KH periodic orbit. The KH periodic orbits also structure the quantum dynamics through its invariant manifolds. However, it is unclear if this organization is sufficient for the wavefunction to be characterised as a KH state.

\section*{Conclusions}
In this article, we have shown that the KH state corresponds to a single periodic orbit with the same period as the laser field, despite the fact that the KH approximation is largely invalid. This KH state exists in a wide range of parameters of the laser and the atom, and in most relevant cases is an elliptic or weakly hyperbolic periodic orbit. We have used the KH approximation as a guide to find it by gradually turning on the neglected terms in the Hamiltonian. As expected, the phase-space picture in the KH approximation bears very little to no resemblance with the true dynamics in the KH coordinates, but it has a useful component as a methodological tool to identify and follow the fate of the main periodic orbits in the integrable case. 
We have shown that the identified periodic orbit has a strong influence on the quantum wavefunction, where the effect of this periodic orbit is clearly visible as a scar lasting several laser cycles if weakly hyperbolic or much longer if elliptic. The existence of this KH state opens up a formidable avenue to exploit these Rydberg-like atoms where the level of excitation is piloted by tuning relevant parameters of the laser field. This could be achieved by controlling the classical KH periodic orbit identified in the present work. 

\begin{acknowledgments} 
The authors would like to thank Fran\c{c}ois Mauger for helpful discussions. 
This research received financial support from the French National Research Agency through Grants No.~ANR-21-CE30-0036-03-ATTOCOM.
JD acknowledges funding from the European Union’s Horizon Europe research and innovation program under the Marie Skłodowska-Curie Grant Agreement No.~101154681. Funded by the European Union.
Views and opinions expressed are however those of the author(s) only and do not necessarily reflect those of the European Union or the European Commission. Neither the European Union nor the European Commission can be held responsible for them.
\end{acknowledgments}

\section*{Author Contribution Statement}
All authors contributed equally to the paper.

\section*{Data Availability Statement}
Data sets generated during the current study are available from the corresponding author on reasonable request.


%

\end{document}